\documentclass[a4paper,twoside,10pt,showpacs,showkeys]{revtex4}

\usepackage[USenglish]{babel} 
\usepackage[T1]{fontenc}
\usepackage[ansinew]{inputenc}
\usepackage{subfigure}
\usepackage{lmodern} 
\usepackage{amsmath}
\usepackage{amstext}
\usepackage{amsopn}
\usepackage{amsfonts}
\usepackage{amssymb}
\usepackage{bbm}
\usepackage{accents}
\usepackage{empheq}
\usepackage{graphicx}
\usepackage{epsf}
\usepackage{graphics}

\usepackage{graphicx} 

\usepackage{amsmath}
\usepackage{amsthm}
\usepackage{amsfonts}
\usepackage{amstext}
\usepackage{empheq}
\usepackage{graphicx}
\usepackage{caption}

\begin{document}



\title{Background gauge renormalization and BRST identities}
\author{J. Frenkel$^{a}$\footnote{Corresponding author}and J C Taylor$^{b}$}
\affiliation{$^a$ Instituto de Fisica, Universidade de Sao Paulo, 05508-090, Sao Paulo. SP,Brazil}
\affiliation{$^b$ Department of Appled Mathematics and Theoretical Physics, University of Cambridge, Cambridge, UK}
\email{jfrenkel@usp.br (J. Frenkel), jct@damtp.cm.ac.uk (J. C. Taylor) }



\pagestyle{plain} 



\def\G{\Gamma}
\begin{abstract}
We show how the BRST identities can be used to control the renormalization of the background gauge in QCD, in spite of the fact that
certain one-particle reducible graphs have to be omitted in general. We obtain the all order renormalized effective action for the background field theory.
\end{abstract}


\maketitle
\section{Introduction}
The background field formalism is  a method to define elements of the effective action $\G$ \cite{dewitt,abbott,weinberg}. The field $A$ is divided into
a quantum part $Q$ and a classical background $B$, so that $A=Q+B$. The action is a function of $A$ only, $\G_0(Q+B)$. No source $J$ is needed at this stage.
Carrying out the Feynman path integral over $Q(x)$ yields the effective action $\G(B)$. However, this gives  a trivial result
unless certain one-particle reducible (1-PR) graphs are omitted. These are graphs resulting from vertices which are linear in $Q$.
(See for example section 16.1 of \cite{weinberg}).

In gauge theories, the method is adapted to the background gauge method. (See \cite{dewitt}, \cite{abbott} and section 17.4 of \cite{weinberg}.) The gauge-fixing term is made to depend upon $B$, for instance
\begin{equation}
L_{GF}=-\frac{1}{2\alpha}[D_{\mu}(B).Q^{\mu}]^2,
\end{equation}
where $\alpha$ is a gauge-fixing parameter and $D(B)$ is the covariant derivative
\begin{equation}
D_{\mu}(B)=\partial_{\mu}+gB_{\mu}\wedge,
\end{equation}
(we suppress colour indices, and use the notation $B_{\mu}.Q_{\nu}=B^a_{\mu}Q^a_{\nu}$, $(B_{\mu}\wedge Q_{\nu})^a=f^{abc}B_{\mu}^b Q_{\nu}^c$ for example).
The result is that $\G(B)$ has gauge invariance under
\begin{equation}
\delta B_{\mu}(x)=D_{\mu}(B)e(x),
\end{equation}
where $e(x)$ is an arbitrary infintesimal classical gauge parameter.

This is an efficient method for calculating the $\beta$-function \cite{abbott,weinberg, capper, mckeon,pickering}, and has also been used in perturbative gravity \cite{thooft,goroff}.

If the background method is used  to two-loop order or higher, the subgraphs are functionals of $Q$ as well as $B$. We may introduce a current $J_{\mu}$ interacting via $J_{\mu}.Q^{\mu}$, thus defining a generating functional $W(B,J)$ for connected graphs; and thence, by a Legendre transform,
an action $\G(B,Q)$. Although 1-PR graphs in general are allowed in $W$, some must be excluded. Examples of excluded graphs in
$W$ and in $\Gamma$ are shown in Fig.1. 

$\G(B,Q)$ has background gauge symmetry under
\begin{equation}
 \delta B_{\mu}=D_{\mu}(B)e,\\\\\ \delta Q_{\mu}=g\hspace{.02cm}
Q_{\mu} \wedge e.
\end{equation}
This symmetry is not in itself sufficient to control the renormalization of $\G(B,Q)$. In addition, BRST
symmetry is called upon. The zeroth order action, $\Gamma^{(0)}(B,Q)$ is BRST invariant, but 
the omission of some 1-PR graphs from $W(B,J)$ (see Fig.1) destroys BRST symmetry. It is this dilemma which we solve. We present no new results,
but we aim to clarify the reasoning behind the use of BRST in the background gauge \cite{grassi,barvinski}.

The usual method to derive the BRST symmetry equation for the effective action starts from the symmetry of $W$. To try to apply this directly in the background formalism, one would have to start from the correct generating function for this formalism, which has some 1-PR graphs excluded, and which we call $\bar{W}(B,J)$. This attempt is then bound to fail, because the BRST symmetry is lost.

We find that there is an effective action $\bar{\G}(B,Q)$ which generates $\bar{W}(B,J)$, which does not have BRST
symmetry but it maintains the background gauge symmetry (3). Nevertheless, the BRST symmetry of $\G(B,Q)$ can be used indirectly
to control the renormalization of $\bar{\G}(B,Q)$, as we show in the next section.

\begin{figure}[h!]
    \centering
    \subfigure[]{\includegraphics[scale=0.5]{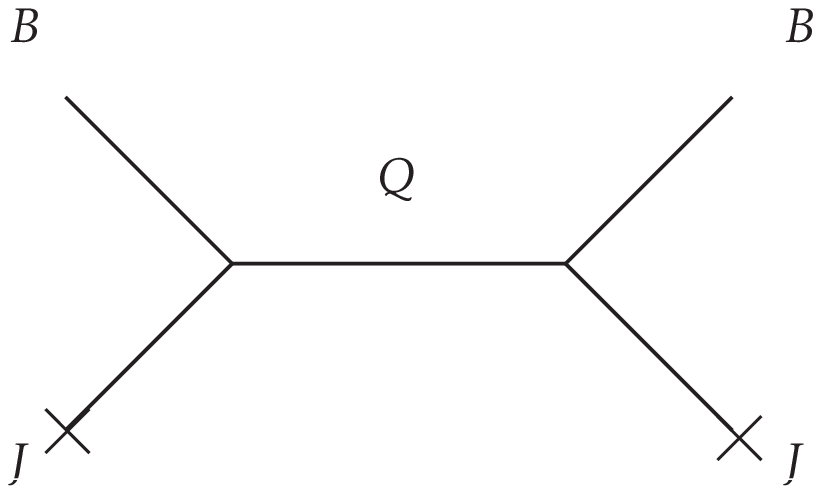}}
    \subfigure[]{\includegraphics[scale=0.5]{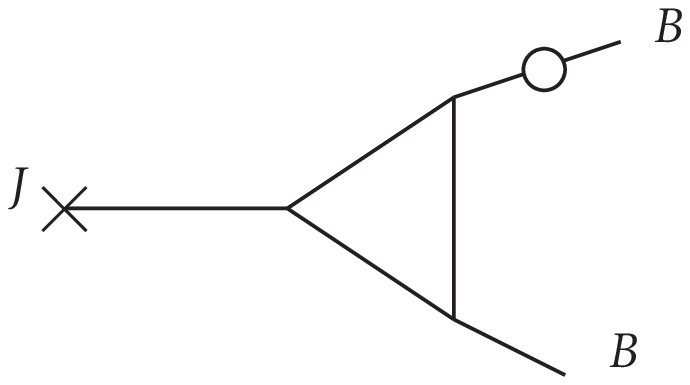}}
	\subfigure[]{\includegraphics[scale=0.6]{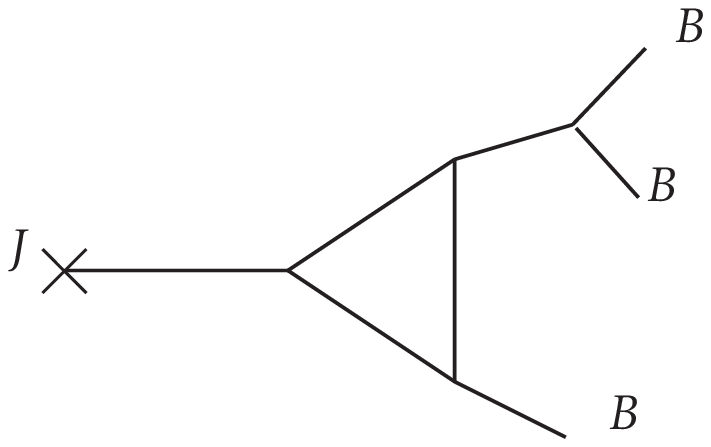}}
	\subfigure[]{       \includegraphics[scale=0.6]{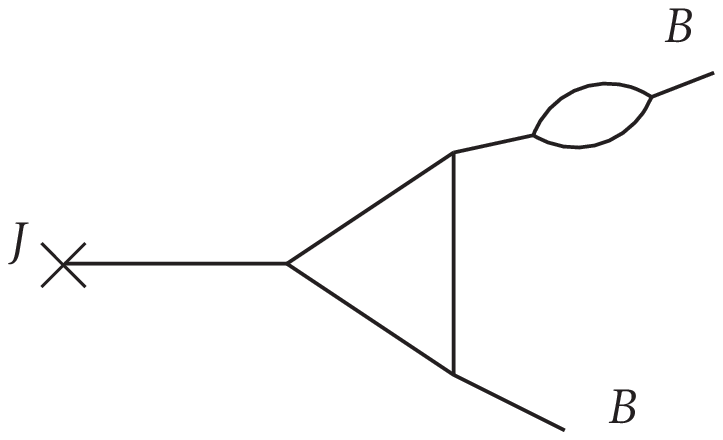}}
	\subfigure[]{\includegraphics[scale=0.6]{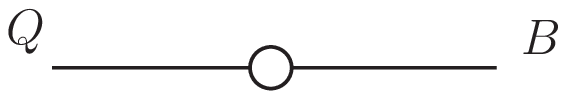}}
	\subfigure[]{\includegraphics[scale=0.7]{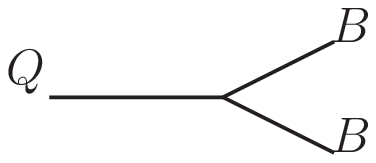}}
	\subfigure[]{\includegraphics[scale=0.7]{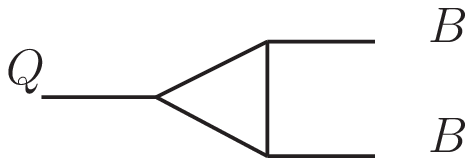}}
   \caption{Examples of graphs in $W(B,J)$ and $\Gamma(B,Q)$. (a) is 1-PR graph in $\bar{W}$. (b), (c), (d) are 1-PR graphs not allowed in $\bar{W}$. (e), (f), (g) are graphs not allowed in $\bar{\Gamma}$ because they would generate forbidden graphs in $\bar{W}$. A cross indicates the action of the source $J$. The little circle stands for the $Q{\tiny-}B$ mixing term in $L$ in equation (6).}
\end{figure}
\section{The effective action $\bar{\G}$}
The zeroth order action is
\begin{equation}
\G^{(0)}(B,Q,c,c*;u,v;g)= \int d^4x L
\end{equation}
where
\begin{equation}
L=-\frac{1}{4}F_{\mu\nu}(B+Q)F^{\mu\nu}(B+Q)-[D_{\mu}(B)c^*].D^{\mu}(B+Q)c+u_{\mu}.D^{\mu}(B+Q)c-\frac{g}{2}v.(c\wedge c),
\end{equation}
where $c,c^*$ are ghost and anti-ghost, and $u$ and $v$ are Zinn-Justin sources, and the gauge-fixing term (1) is omitted. (For simplicity, we leave out background ghost fields, and we leave out quarks altogether. For the general case, see section 17.4 of \cite{weinberg}.) 
In (6), $F$ is defined by $gF_{\mu\nu}(B)=[D_\mu(B), D_\nu(B)].$
 $L$ is invariant under the BRST tranformations
\begin{equation}
sB=0,\;\\\ sQ_{\mu}=D_{\mu}(B+Q)c\,\zeta,\;\\\ sc=-\frac{g}{2}c\wedge c\, \zeta,
\end{equation}
($\zeta$ being an infintesimal anti-commuting constant) and hence $\G^{(0)}$ obeys the BRST equations
\begin{equation}
\int d^4x \left[\frac{\delta \G^{(0
)}}{\delta Q_{\mu}(x)}.\frac{\delta \G^{(0)}}{\delta u^{\mu}(x)}+\frac{\delta \G^{(0)}}{\delta c(x)}.\frac{\delta \G^{(0)}}{\delta v(x)}\right]=0,
\end{equation}
\begin{equation}
 \frac{\delta \Gamma^{(0)}}{\delta c^*}+D_{\mu}(B)\frac{\delta \Gamma^{(0)}}{\delta u_{\mu}}=0.
\end{equation}

It follows by the usual derivation (using first the invariance properties of $W(B,J)$; see, for example, chapter 12 in \cite{taylor} and section 12-4 in \cite{itzykson}) that the  effective action $\G$ to all orders
satisfies
\begin{equation}
\int d^4x \left[\frac{\delta \G}{\delta Q_{\mu}(x)}.\frac{\delta \G}{\delta u^{\mu}(x)}+\frac{\delta \G}{\delta c(x)}.\frac{\delta \G}{\delta v(x)}\right]=0,
\end{equation}
\begin{equation}
 \frac{\delta \Gamma}{\delta c^*}+D_{\mu}(B)\frac{\delta \Gamma}{\delta u_{\mu}}=0.
\end{equation}
The identities resulting from (10) are different from the usual ones for gauge theories, because of the dependence on the extra field $B(x)$. We have explicitly verified,  by calculations to order
$g^3$, some examples of these identities.

But unfortunately $\G$ is not the correct effective action for the background method because it leads to unwanted 1-PR graphs in $W(B,J)$. (See Fig.1, for examples).
These can be removed at one-loop order by replacing $\G^{(0)}$ by $\bar{\G}^{(0)}$ defined by the operation
\begin{equation}
\bar{\G}^{(0)}= \Xi (\G^{(0)}) \equiv \G^{(0)}-\int d^4x \left[Q^{\mu}(x).\left\{\frac{\delta \G^{(0)}}{\delta Q^{\mu}(x)}\right\}_{Q=c=0}\right].
\end{equation}
To higher order, terms may appear in $\G$ which again generate forbidden graphs in $W(B,J)$; so to all orders we must replace $\G$
by the effective action $\bar{\G}$ defined analagously to (12) by
\begin{equation}
\bar{\G}=\Xi \hspace{0.02cm}(\G).
\end{equation}
The operation $\Xi$ maintains background gauge invariance (4).
However, $\bar{\G}^{(0)}$ is not invariant under the BRST transformation (7). This is evident because the subtracted term in (12) is
\begin{equation}
\int d^4x Q^{\mu}(x).[D^{\nu}(B)F_{\mu\nu}(B)].
\end{equation}
It follows that $\bar{\G}$ does not satisfy the BRST equations. This seems to pose a difficulty in applying BRST
in the background method.

We propose the following solution to this problem. First we deduce, by the standard method using BRST, the renormalized form of $\G$. Call it
$\G_R$. Suppose we renormalize by iteration, first removing the divergences up to and including $n-1$ loops, and then
looking for the divergences at $n$ loops. Let $\G^{(n)}$ be the $n$-loop action. Then, taking the divergent parts in (13),
\begin{equation}
\bar{\G}^{(n)}_{div}=\Xi (\G^{(n)}_{div}).
\end{equation}
It follows that the counter-terms necessary to cancel the divergences obey an equation of the same form
\begin{equation}
\bar{\G}^{(n)}_{CT}=\Xi( \G^{(n)}_{CT}),
\end{equation}
and thus to all orders,
\begin{equation}
\bar{\G}_R = \Xi(\G_R).
\end{equation}
Thus we may apply BRST to renormalize $\G$ by the standard methods, and deduce the renormalization of $\bar{\G}$ by the application
of the operation $\Xi$ defined by (12).

\section{Renormalization}
As explained in the last section, we study first the renormalization of $\G$, although this is not the action for the background gauge. 
Both the background symmetry as well as the BRST invariance are necessary to fix the counter-terms which cancel the divergences in $\G$ \cite{grassi,barvinski}.
(The background gauge symmetry under (4) has to be completed to include the fields $c$, $c^*$, $u$, $v$, 
each of which transforms covariantly in the same way as $Q$.)
There is a complication in the background gauge because the second BRST equation (11) depends explicitly on $g$, so it is not obvious that
$g$ is renormalized just by scaling. This problem is resolved if $B$ is simultaneously rescaled. Apart from these considerations, renormalization is similar to that in the ordinary gauge theory, as discussed for instance in section 12-4-3 of \cite{itzykson}. 
The BRST equations (10), (11) and the background gauge symmetry are all preserved by the rescalings
\begin{equation}
g\rightarrow  Z_g g,\; B\rightarrow Z_g^{-1}B, \; Q\rightarrow Z_Q^{1/2}Q,\;c\rightarrow \tilde{Z}^{1/2}c,\; c^* \rightarrow  \tilde{Z}^{1/2}c^*,\; u\rightarrow \tilde{Z}^{1/2}u,\;  \;v\rightarrow Z_Q^{1/2}v.
\end{equation}
Under these rescalings, each of the two terms on the left of the BRST equation (10) acquires the same factor $(Z_Q \tilde{Z})^{-1/2}$. (This is the conventional choice, but it is not unique because any value for this common factor would be possible.).
A mixing of $B$ into $Q$ is forbidden by the second equation of (4).
By dimensions, Lorentz invariance and rigid gauge-invariance, the renormalized action $\G_R$ must be similar to  ${\G}^{(0)}$
in (5) and (6), but with the allowed scalings (18), so it must have the form 
	\begin{equation}
	\G_R=\G^{(0)}(Z_g^{-1}B,Z_Q^{1/2}Q,  \tilde{Z}^{1/2}c, \tilde{Z}^{1/2}c^*,\tilde{Z}^{1/2}u,    Z_Q^{1/2}v ; Z_g g).
	\end{equation}

	 Finally, the renormalized action for the background field theory is given, according to (17), by
	\begin{equation}
	\bar{\G}_R= \Xi (\G_R),
	\end{equation}
	where the terms in (19) which are linear in $Q$, but independent of $c$, have been subtracted.
	
	\section{Conclusion}
	We have shown how to renormalize to all orders the Yang-Mills theory in the background gauge, in a way consistent with BRST and background gauge invariance. Our method also allows for the omission of 1-PR graphs in the background method, which leads to the renormalized effective action (20).
	\\
	\\
	J.F. would like to thank A.L.M. Britto for helpful conversations, and CNPq (Brazil) for a grant.

\end{document}